\documentstyle[12pt]{article}

\begin{document}

\baselineskip=0.8cm
{\large
\begin{center}
REAL SPACE APPROACH TO\\
\vspace{0.15cm}
ELECTRONIC-STRUCTURE CALCULATIONS \\
\end{center}
}
\begin{center}
E\hspace{0.4mm}i\hspace{0.4mm}j\hspace{0.4mm}i  Tsuchida and Masaru Tsukada \\
\end{center}
\begin{center}
 Department of Physics, Faculty of Science, University of Tokyo, \\
 Hongo 7-3-1, Bunkyo-ku, Tokyo 113, Japan \\
\end{center}

\begin{abstract}
\baselineskip=0.8cm
We have applied the {\it Finite Element Method} to the self-consistent
electronic structure calculations of molecules and solids
for the first time. In this approach all the calculations
are performed in "real space" and the use of non-uniform mesh is made possible,
 thus enabling us to deal with localized systems with ease.
To illustrate the utility of this method, we perform an all-electron
calculation of hydrogen molecule in a supercell with LDA approximation.
Our method is also applicable to mesoscopic systems.\\
\ \\
\end{abstract}

\newpage

\begin{center}
{\large\bf I. Introduction} \\
\end{center}

\ \ \ Electronic-structure calculations of materials have a long history and
various methods have been developed for that purpose.
One of the most popular methods for electronic-structure calculations
 of condensed matter systems today is the use of plane-wave basis sets and
pseudo-potentials (PWPP) with
local density approximation (LDA) \cite{Payne}.
This method usually has enough accuracy for
 quantitative discussion, and reasonable efficiency to treat large
systems. However, for localized systems such as the first row elements or
 transition metals, the use of PWPP leads to a large number of plane-waves,
 thus making the calculation almost impossible.

One way to avoid this difficulty is to use the ultrasoft pseudo-potentials
 \cite{VDB}. In this method, the constraint of norm-conservation is
 relaxed, and the number of plane-waves necessary for the calculation
is considerably reduced. However, the deficit charge must be
 restored at every step, and the calculation becomes complicated.
Furthermore, for systems with large vacuum regions such as molecules or
 slab model of surfaces in supercell geometry, even the ultrasoft
pseudo-potential does not significantly reduce the number of
plane-waves.
 So an efficient method to deal with such systems is desirable.

\begin{center}
{\large\bf II. Details of the Calculation} \\
\end{center}

\ \ \ Here we propose another way to solve above difficulties.
In our method, all the calculations are performed in real space,
and non-uniform meshes are used. First, the unit cell is divided into
ortho-rhombic elements in real space, and the values of the wavefunctions
 on each vertex of the elements are taken as the basic variables.
Then the wavefunctions are interpolated in each element as follows.

\begin{equation}
\left. \psi (x,y,z) \right|_{element}= \sum_{s=1}^8 \psi_s \cdot \omega_s,
\end{equation}

\ \\
\begin{equation}
\omega_1 = \left( 1 - \frac{x}{h_1} \right)
              \left( 1 - \frac{y}{h_2} \right)
              \left( 1 - \frac{z}{h_3} \right),
\end{equation}
\begin{equation}
\omega_2 = \frac{x}{h_1}
              \left( 1 - \frac{y}{h_2} \right)
              \left( 1 - \frac{z}{h_3} \right),
\end{equation}
\begin{equation}
\omega_3 = \frac{y}{h_2}
              \left( 1 - \frac{x}{h_1} \right)
              \left( 1 - \frac{z}{h_3} \right),
\end{equation}
\begin{equation}
\omega_5 = \frac{z}{h_3}
              \left( 1 - \frac{x}{h_1} \right)
              \left( 1 - \frac{y}{h_2} \right),
\end{equation}
\begin{equation}
\omega_4 = \frac{x}{h_1}
              \frac{y}{h_2}
              \left( 1 - \frac{z}{h_3} \right),
\end{equation}
\begin{equation}
\omega_6 = \frac{x}{h_1}
              \frac{z}{h_3}
              \left( 1 - \frac{y}{h_2} \right),
\end{equation}
\begin{equation}
\omega_7 = \frac{y}{h_2}
              \frac{z}{h_3}
              \left( 1 - \frac{x}{h_1} \right),
\end{equation}
\begin{equation}
\omega_8 = \frac{x}{h_1}
              \frac{y}{h_2}
              \frac{z}{h_3},
\end{equation}
where $ h_1, h_2, h_3 $ denote the length of the three sides of the element
 and $ \psi_1, \cdots , \psi_8 $ denote the value of the wavefunction
 on each vertex of the element (Fig.1).
 Note that $ \omega_s $ equals 1 at the {\it s}-th
 vertex and 0 at the other vertices.

In this way, the wavefunctions are approximated by continuous
piecewise polynomial functions.
Here we put the values of the wavefunction of the {\it i}-th band
on every vertex in the unit cell in some appropriate order and write it as
$ \,\vec \psi_i $.

The density corresponding to the wavefunction is approximated in the
same way and represented by $ \vec n $. Here the values of the density
is calculated from the square sum of the wavefunctions. In order to
conserve the total charge, re-normalization of the density is
performed. Though the present treatment of the density is not
satisfactory, it causes no serious troubles such as instabilities,
so we adopt it for the time being.

In this approach, the total energy of the system is represented in
the following way. First, we confine ourselves to the periodic systems
 to handle the long range Hartree potential efficiently.
Furthermore, for simplicity we consider only real wavefunctions and
bare coulomb potentials, though the extension to complex wavefunctions
and separable norm-conserving pseudo-potentials is straightforward.

The total energy per unit cell is given by
\begin{equation}
E_{total} = E_{kin} + E_{pot} + E_{hartree} + E_{xc} + E_{ewald}
\end{equation}
in the LDA approximation. Rydberg units are used th\-rou\-ghout the paper.

The explicit form of the kinetic energy is
\begin{equation}
E_{kin} = \sum_i \int_{cell} \left| \nabla \psi_i \right|^2 \, d \vec r .
\end{equation}

This integral can be calculated exactly in each element for the interpolated
expression, and the result is the quadratic form
of the values of wavefunctions at the vertices. So, with the aid of a
constant matrix $ G $, we can write
\begin{equation}
E_{kin} = \sum_i {}^t \vec \psi_i \cdot G \cdot \vec \psi_i.
\end{equation}

We now go on to the potential energy. After the divergences
of the long range Coulomb potentials of nuclei and electrons
are cancelled, the potential energy is expressed as
\begin{equation}
E_{pot} = -8 \pi \sum_{\vec G \ne 0} \frac{n\,(\vec G)}{G^2} \cdot
\left(\sum_i Z_i \exp{(i\, \vec G \cdot \vec r_i)} \right) ,
\end{equation}
where $ n\,(\vec G) $ is the reciprocal-space representation of the
density and $ Z_i $ and $ \vec r_i $ denote the charge and the position of
each nucleus in the unit cell respectively.
Here we define a potential in real space as
\begin{equation}
V\,(\vec r) = - \frac{8 \pi}{V_{cell}} \sum_i Z_i \,
\left( \sum_{\vec G \ne 0} \frac{1}{G^2}\,\exp{(i\, \vec G \cdot
(\vec r - \vec r_i)) } \right) .
\end{equation}
In practice, the above sum is calculated with
{\it Ewald's method} \cite{ZM}.
Using this potential, the potential energy can be written as
\begin{equation}
E_{pot} = \int_{cell} V\,(\vec r) \, n\,(\vec r) \,d \vec r.
\end{equation}
This can again be calculated in each element, and the result is
the linear form of the density. So we can write with the use of a
constant vector
$ \vec v $,
\begin{equation}
E_{pot} =  {}^t \vec v \cdot \vec n,
\end{equation}
where each element of $ \vec v $ is calculated as the sum of several
integrals of the form
$ \int_{element} \,V\,(\vec r)\, \omega_s \,(\vec r) \, d \vec r $.

In the same way, the Hartree energy can be written as
\begin{equation}
E_{hartree} = 4 \pi \int_{cell} \varphi \, (\vec r) \, n \, (\vec r) \,
d \vec r ,
\end{equation}
with
\begin{equation}
\label{vp}
\varphi \, (\vec r) = \sum_{\vec G \ne 0} \frac{n\,(\vec G)}{G^2}
\, \exp{(i \, \vec G \cdot \vec r)} .
\end{equation}

Note that $ \varphi \, (\vec r) $ depends on the desity in contrast
with $ V\, (\vec r) $.
As a result, $ \varphi \, (\vec r ) $ must be calculated every time
the density is changed.
The direct use of the right hand side of (\ref{vp})
 is costly, because the fast Fourier transform (FFT) is necessary
to obtain the values of $ n\,(\vec G) $  for all $ \vec G $.
So we calculate $ \varphi \, (\vec r) $ by minimizing
\begin{equation}
\label{Ip}
I \,\left[\varphi \right] \equiv \int_{cell} \left\{ \frac12
\left| \nabla \varphi \right| ^2 - (n\,(\vec r) - \bar n) \cdot
\varphi\,(\vec r) \right\} d \vec r
\end{equation}
with the conjugate gradient method under the contraint,
\begin{equation}
\int_{cell} \varphi\,(\vec r) \, d \vec r  = 0 ,
\end{equation}
where $ \bar n $ denote the average of the density.
It can be easily shown that $ \varphi\,(\vec r) $ given in
Eq.(\ref {vp}) minimizes the functional (\ref {Ip}) under the
same constraint.
This technique for calculating Hartree potential is adopted in \cite{DCAJ}
for a different reason.
Here we approximate $ \varphi\,(\vec r ) $ in the same way as the density,
and write it as $ \vec \varphi $ .
In this method,
$ \vec \varphi $ need not be so exact when the total energy
is far from convergent.
Accordingly, when the density is changed, only a few iterations are
 necessary to get the new $ \vec \varphi $ .
Similar idea on this procedure is described in \cite{CP2}.

Now we can calculate the Hartree energy and the result is the
bilinear form of $ \vec n $ and $ \vec \varphi $ . Thus we can write
 with a constant matrix $ F $ ,
\begin{equation}
E_{hartree} = 4 \pi \, {}^t \vec \varphi \cdot F \cdot \vec n.
\end{equation}

The rest of the total energy, $ E_{xc} $ and $ E_{ewald} $ can be
easily handled, and will not be given here.
\ \\

Now that the total energy is expressed as the function of \
$ \{ \vec \psi_i \} $ ,
we go on to minimize it with the conjugate gradient method \cite{Payne}
under constraints
\begin{equation}
\int \psi_i (\vec r ) \, \psi_j (\vec r) \, d \vec r = \delta_{ij},
\end{equation}
or in our notation,
\begin{equation}
{}^t \vec \psi_i \cdot F \cdot \vec \psi_j = \delta_{ij}.
\end{equation}

Here $ F $ is a sparse and symmetric matrix whose diagonal element
is proportional to the volume of the element to which the vertex
belongs. So if we apply the conjugate gradient method directly when
we are working with non-uniform meshes,
the conjugate gradient vector is destroyed in the orthogonalization
procedure, and the minimum of the total energy is never reached.
In order to avoid this difficulty, we transform the wavefunctions as
\begin{equation}
\vec \psi_i' = T \cdot \vec \psi_i \;\;\;\;\; \mbox{for} \; \forall \, i.
\end{equation}
Then the constraints become
\begin{equation}
{}^t \vec \psi_i' \cdot F' \cdot \vec \psi_j' = \delta_{ij},
\end{equation}
with
\begin{equation}
F' = {}^t \left( T^{-1} \right) \cdot F \cdot \left( T^{-1} \right).
\end{equation}
We choose the matrix $ T $ that makes $ F' $ nearly the unit matrix.
Another possible solution is to use the {\it unconstrained minimization}
\cite{MGC},\ in which no explicit orthogonalization is required.
Note that this transformation is different from the usual
pre-conditioning, which is determined by the form of the total energy.

\begin{center}
{\large\bf III. Example}\\
\end{center}

\ \ \ To demonstrate the utility of the present method, we calculated the
equilibrium structure of hydrogen mo\-lecule in a cubic supercell
as illustrated in Fig.2.
An all-electron calculation was performed with non-uniform meshes shown
in Fig.3.
At the boundary between elements with different size, the wavefunctions
 are constrained to be continuous.
The calculated density is shown in Fig.4.
The values of the calculated bond length and vibrational frequency
are given in Table I and compared with those from experiments and
 other theories. The agreement is satisfactory.

\ \\
\begin{tabular}{lcc}
\hline
\hline
              & Bond Length (a.u.) & Vibrational Frequency ($cm^{-1}$) \\
\hline
 This work    &              1.46  &                        4424   \\
 Other theory &              1.45  &                        4277   \\
 Experiment   &              1.40  &                        4400   \\
\hline
\hline
\end{tabular}
\ \\

Table I. \ The other theory is from Ref \cite{PA}, in which
LDA calculations are performed with Gaussian orbitals.
The same form for $ \epsilon_{xc} $ is adopted in these calculations
 \cite{CA,VWN}.\\

\begin{center}
{\large\bf IV. Discussion} \\
\end{center}

\ \ \ The advantages of this method are as follows.
(i) Since all calculations are performed in real space,
no FFT is necessary in contrast with PWPP.
(ii) We can use non-uniform meshes. In the terminology of PWPP,
 we can effectively change the cutoff-energy {\it locally}.
 Accordingly, we can express localized orbitals without any serious
difficulties. Furthermore, we can reduce the number of variables considerably
when we calculate the electronic-structure of surfaces
in a supercell configuration as illustrated in Fig.5.
 Similar idea is introduced in Ref \cite{DCAJ,GY2}, in which plane-wave is
used with adative Riemannian metric.
(iii) In this method, the effect of the potential of the
nucleus appears only in an integral form,
so the divergence of the bare coulomb potential at the origin
does not cause any difficulty.
(iv) Orbitals localized in given regions of space
\cite{GP} and their overlap matrix are naturally treated in this approach.
(v) Our method is much easier to implement than PWPP.

One clear limitation of this method is that it will be ineffective
if we perform dynamical simulations with non-uniform meshes, which lead to
 the Pulay-like forces \cite{PF}.

\begin{center}
{\large\bf V. Conclusion} \\
\end{center}

\ \ \ In conclusion, we have applied the {\it Finite Element Method} to
the self-consistent calculation of electrons successfully for the
first time. Since this method has several advantages over PWPP,
it will be considerably favorable for the calculations of some systems.
Extensions of this method to higher order interpolation
formulas and separable pseudo-potentials are under way.\\
\ \\
Acknowledgement\ --\
One of the authors (E.Tsuchida) would like to thank Katsuyoshi Kobayashi
and Jun Yamauchi for fruitful discussions, and Shigenobu Kimura for
providing useful subroutines.
The numerical calculations were performed on HP-9000.

\newpage

\ \\
\ \\
\ \\
Fig.1. \ \ An element and its vertices.\\
\ \\
Fig.2. \ Hydrogen molecule in a cubic supercell (12 a.u. on a side).
The calculations were performed with an eighth of the supercell.\\
\ \\
Fig.3. \ \ In practice, we used the mesh twice as dense as this
figure. The mesh is taken approximately logarithmic. \\
\ \\
Fig.4. \ The calculated density of hydrogen molecule. The singularity
at the nucleus is well described.\\
\ \\
Fig.5. \ In the vacuum region, only a small number of elements will
suffice.\\
\ \\

\end{document}